\documentclass[english,aps,onecolumn,amsmath,amssymb,showpacs,superscriptaddress,notitlepage,longbibliography]{revtex4-1}

\usepackage{amsmath,amsfonts,amssymb}
\usepackage{graphicx}
\usepackage[colorlinks=true, allcolors=blue]{hyperref}

\newcommand{\mat}[1]{\underline{\underline{#1}}}
\newcommand{\vc}[1]{\mathbf{#1}}

\begin{document} 

\title{Density-functional description of materials for topological qubits and superconducting spintronics}

\author{Philipp Rüßmann}
\affiliation{Institute for Theoretical Physics and Astrophysics, University of Würzburg, 97074 Würzburg, Germany}
\affiliation{Peter Grünberg Institut and Institute for Advanced Simulation, Forschungszentrum Jülich and JARA, 52425 Jülich, Germany}

\author{David Antognini Silva}
\affiliation{Peter Grünberg Institut and Institute for Advanced Simulation, Forschungszentrum Jülich and JARA, 52425 Jülich, Germany}
\affiliation{Department of Physics, RWTH Aachen University, 52056 Aachen, Germany}

\author{Mohammad Hemmati}
\affiliation{Peter Grünberg Institut and Institute for Advanced Simulation, Forschungszentrum Jülich and JARA, 52425 Jülich, Germany}
\affiliation{Department of Physics, RWTH Aachen University, 52056 Aachen, Germany}

\author{Ilias Klepetsanis}
\affiliation{Peter Grünberg Institut and Institute for Advanced Simulation, Forschungszentrum Jülich and JARA, 52425 Jülich, Germany}
\affiliation{Department of Physics, National and Kapodistrian University of Athens, Panepistimioupolis 15784 Athens, Greece}

\author{Björn Trauzettel}
\affiliation{Institute for Theoretical Physics and Astrophysics, University of Würzburg, 97074 Würzburg, Germany}
\affiliation{Würzburg-Dresden Cluster of Excellence ct.qmat, Germany}

\author{Phivos Mavropoulos}
\affiliation{Department of Physics, National and Kapodistrian University of Athens, Panepistimioupolis 15784 Athens, Greece}

\author{Stefan Blügel}
\affiliation{Peter Grünberg Institut and Institute for Advanced Simulation, Forschungszentrum Jülich and JARA, 52425 Jülich, Germany}

\begin{abstract}
    Interfacing superconductors with magnetic or topological materials offers a playground where novel phenomena like topological superconductivity, Majorana zero modes, or superconducting spintronics are emerging. In this work, we discuss recent developments in the Kohn-Sham Bogoliubov-de Gennes method, which allows to perform material-specific simulations of complex superconducting heterostructures on the basis of density functional theory. As a model system we study magnetically-doped Pb. In our analysis we focus on the interplay of magnetism and superconductivity. This combination leads to Yu-Shiba-Rusinov (YSR) in-gap bound states at magnetic defects and the breakdown of superconductivity at larger impurity concentrations. Moreover, the influence of spin-orbit coupling and on orbital splitting of YSR states as well as the appearance of a triplet component in the order parameter is discussed. These effects can be exploited in S/F/S-type devices (S=superconductor, F=ferromagnet) in the field of superconducting spintronics.
\end{abstract}

\maketitle

\section{INTRODUCTION}
\label{sec:intro}  

Systems where superconductivity meets magnetism or topological materials are a rapidly emerging field. This is not only due to the advent of topological superconductivity, where Majorana zero modes can form building blocks for topological quantum computing~\cite{Kitaev2001, NadjPerge2014}, but it also has become of interest in the field of superconducting spintronics~\cite{Eschrig2015, Cai2023, Melnikov2022}. Novel states of matter generally can exist in complex heterostructures of, for instance, superconductors and magnetic materials, chains of magnetic atoms on superconductors, or at interfaces between superconductors and topological insulators. While a plethora of different experimental techniques allows gaining invaluable insights and simple models can explain general trends, a material-specific theory allows for predictive modeling of complex material combinations on the atomic scale. The Kohn-Sham Bogoliubov-de Gennes (KS-BdG) method provides this capability, which offers new ways to understand the physics of complex superconducting heterostructures and enables pathways toward computational material optimization. Next to experiments and theory, computational physics forms the third pillar of modern physics that is indispensable in overcoming the immense material challenges in the quickly developing field around inhomogeneous superconducting materials. 

Density-functional theory (DFT) for superconductors has been formulated since the works of Oliveira, Gross and Kohn~\cite{ogk-dftsc}. In the spirit of the Bogoliubov-de Gennes approach, the theory extends the single-particle picture of DFT to an electron-hole space with a coupling that, for conventional $s$-wave superconductors, depends on the electron-phonon interaction. This method is typically referred to as the Kohn-Sham Bogoliubov-de Gennes method (KS-BdG). There are two different approaches in the DFT of superconductivity. The first approach is to calculate the electron-phonon coupling entirely from first principles, leading to a fully ab-initio prediction of superconductors \cite{Lueders2005, Profeta2006, Floris2007}. The second approach, analogous to the LDA$+U$ approach to strongly correlated systems in DFT, is to relate the coupling strength to an empirical parameter $\lambda$, which is set by fitting the results to some experimentally accessible physical quantity, e.g.\ the superconducting gap~\cite{Csire2015, Tombulk, Ruessmann2022a}. While the first approach is realistic in systems with only few atoms in the unit cell, it becomes numerically very expensive in systems that require large computational unit cells. This is particularly the case for disordered systems where the second approach is favorable as a compromise between numerical demand and flexibility. It was successfully used to study conventional $s$-wave superconductors~\cite{Csire2015, CsireSchoenecker2016, Tombulk, Ruessmann2022a}, heterostructures of $s$-wave superconductors and non-superconductors~\cite{Csire2016, Csireheterostruc2016, Park2020, Ruessmann2022b, Ruessmann2023}, or impurities embedded into superconductors~\cite{Tomimp, Nyari2021, Saunderson2022, Wu2023}.

In this work we focus on recent developments in the KS-BdG method using the parameter-based approach. This was combined with the Korringa-Kohn-Rostoker Green function formalism (KKR) based on multiple-scattering theory~\cite{Ebert2011}, founded on the seminal work of Suvasini \textit{et al.}~\cite{Suvasini1993} Recently, the KS-BdG approach was also implemented into the open-source JuKKR code~\cite{jukkr}, that allows a full-potential relativistic description of complex materials~\cite{Ruessmann2022a}. As a prime example for the capabilities of the KS-BdG method we choose to focus our attention on superconducting Pb doped with magnetic transition-metal defects. Lead is a well studied elemental superconductor that has low enough electron density so that the $3d$ elements Ti, V, Cr, Mn, Fe, and Co are magnetic when diluted in it. 
We demonstrate two approaches to describe magnetic impurities in superconducting Pb. On the one hand we introduce the ab-initio impurity embedding scheme that allows to describe impurities in the dilute limit as isolated defects in an otherwise perfectly ordered host crystal. On the other hand, we employ the Coherent Potential Approximation (CPA) to deal with finite concentrations of disorder. Both approaches are unique features of a Green's function method~\cite{Dederichs1991}.

Using these two methods we are able to study the appearance of Yu-Shiba-Rusinov (YSR) states \cite{Yu1965, Shiba1968, Rusinov1969} that appear as bound states inside the superconducting gap of Pb~\cite{Ruby2016}. With increasing concentration we find a broadening of these states with the formation of impurity bands up to a critical concentration where superconductivity breaks down. This critical concentration depends on the properties of the respective transition metal impurity. Furthermore, we comment on the appearance of a triplet order parameter that can survive over long distances in superconductor/ferromagnet (SF) junctions. Triplet Cooper pairs can potentially be exploited in SFS-type devices in the field of superconducting spintronics. Overall these examples demonstrate the merit of KS-BdG calculations for superconducting spintronics and quantum information technology.

This work is structured as follows. First, in Sec.~\ref{sec:methods} the KS-BdG method within the framework of KKR is reviewed, and the ab-initio impurity embedding and CPA approaches to disorder embedded in superconductors is introduced. Then, the capabilities of KS-BdG simulations are demonstrated at the example of magnetically-doped Pb in Sec.~\ref{sec:results}. Finally, in Sec.~\ref{sec:discussion} the merit of KS-BdG simulations in the context of materials science involving superconductors, especially in the fields of  topological qubits and superconducting spintronics, is discussed.

\section{METHODS}
\label{sec:methods}

\subsection{The Kohn-Sham Bogolibov-de Gennes method within KKR}
\label{sec:KSBdG}

The Bogoliubov-de Gennes (BdG) method is a powerful theoretical tool used for a microscopic description of superconductors. The advantages of the BdG method are the possibility to describe the excitation spectrum of superconductors with the ability of dealing with inhomogeneous systems, defects in superconductors and complex heterostructures~\cite{BdGbook}. In the context of DFT, the KS-BdG Hamiltonian can be written as~\cite{Oliveira1988, Suvasini1993, Csire2015, Ruessmann2022a}
\begin{equation}
	H^\mathrm{KS}_{\mathrm{BdG}}(\mathbf{x}) = \left(
	\begin{array}{cc}
		H_0^\mathrm{KS}(\mathbf{x})-E_{\mathrm{F}} & \Delta_{\mathrm{eff}}(\mathbf{x}) \\
		\Delta^*_{\mathrm{eff}}(\mathbf{x}) & E_\mathrm{F}-\bigl(H_0^\mathrm{KS}(\mathbf{x})\bigr)^\dagger
	\end{array}
	\right),
	\label{eq:HKSBdG}
\end{equation}
where $E_\mathrm{F}$ is the Fermi energy. Here, we have introduced the normal state Hamiltonian
\begin{equation}
    H^\mathrm{KS}_0(\mathbf{x}) = -\nabla^2 + V_{\mathrm{eff},0}(\mathbf{x}) + \mathbf{B}_{\mathrm{eff}}(\mathbf{x})\cdot \boldsymbol{\sigma},
    \label{eq:H0}
\end{equation}
which is the conventional Kohn-Sham Hamiltonian describing the normal state (Rydberg atomic units are used where $\hbar=1$), as well as the effective superconducting pairing potential $\Delta_\mathrm{eff}$. Note that the potential term in Eq.~\eqref{eq:H0} consists of a non-magnetic  ($V_{\mathrm{eff},0}$) and a magnetic part of the potential (last term on r.h.s.), where $\boldsymbol{\sigma}$ denotes the vector of Pauli matrices.
The effective single-particle potentials in Eq.~\eqref{eq:HKSBdG} are functionals of the charge density $\rho(\mathbf{x})$, the magnetization density, $\boldsymbol{\mu}(\vc{x})$, and the anomalous density $\chi(\mathbf{x})$ (which is the superconducting order parameter)~\cite{Oliveira1988, Suvasini1993},
\begin{eqnarray}
	V_{\mathrm{eff},0}(\mathbf{x}) &=& V_{\mathrm{ext}}(\mathbf{x}) + 2 \int \frac{\rho(\mathbf{x}')}{|\mathbf{x}-\mathbf{x}'|} \mathrm{d}\mathbf{x}' + \frac{\delta E_\mathrm{xc}[\rho,\boldsymbol{\mu},\chi]}{\delta \rho(\mathbf{x})}, \label{eq:Veff} \\ 
	\vc{B}_{\mathrm{eff}}(\mathbf{x}) &=& \vc{B}_{\mathrm{ext}}(\mathbf{x}) + \frac{\delta E_\mathrm{xc}[\rho,\boldsymbol{\mu}, \chi]}{\delta \boldsymbol{\mu}(\mathbf{x})}, \label{eq:Beff} \\ 
	\Delta_{\mathrm{eff}}(\mathbf{x}) &=& \lambda_i \chi(\mathbf{x}). \label{eq:Deff}
\end{eqnarray}
For the functional derivatives of the exchange correlation functional $E_\mathrm{xc}$, the approximations used routinely in DFT can be used. This simplifies the last term in Eq.~\eqref{eq:Veff} to the standard formulation of the local density or generalized gradient approximations (LDA, GGA). The form of the pairing potential is an approximation introduced by Suvasini \textit{et al.}~\cite{Suvasini1993}, who introduced the set of effective coupling constants $\lambda_i$. The different $\lambda_i$  are allowed to vary in the different sites $i$ of a computational unit cell which enables the description of inhomogeneous systems consisting of superconductors and non-superconductors such as magnets~\cite{Tomimp, Wu2023}, normal metals~\cite{Csireheterostruc2016,Csire2016,Csire2018, Ruessmann2023}, or topological insulators~\cite{Park2020,Ruessmann2022b}.

In the KKR formalism it is straight-forward to include the BdG formalism where using multiple-scattering theory the Green's function for the KS-BdG Hamiltonian, Eq.~\eqref{eq:HKSBdG}, is found~\cite{Suvasini1993, Csire2015}.
This results in a generalized structure of the Green's function ($\vc{x}$ is the position vector and $E$ the energy)
\begin{equation}
    \mat{G}(\vc{x}, \vc{x'}; E) = \left(
    \begin{array}{cc}
        G^{e,e}(\vc{x}, \vc{x'}; E) & G^{e,h}(\vc{x}, \vc{x'}; E) \\
        G^{h,e}(\vc{x}, \vc{x'}; E) & G^{h,h}(\vc{x}, \vc{x'}; E)
    \end{array}
    \right),
    \label{eq:GFBdG}
\end{equation}
that consists of four particle-hole blocks that enter in the KS-BdG Hamiltonian of Eq.~\eqref{eq:HKSBdG}. Analogous to the conventional KKR formalism describing the electronic structure in the normal state, the charge $\rho$, magnetization $\boldsymbol{\mu}$, and anomalous density $\chi$ can be computed from the (off-)diagonal blocks of the Green's function~\cite{Suvasini1993, Csire2015}
\begin{eqnarray}
    \rho &=& -\frac{1}{\pi} \int_{-\infty}^{\infty} \mathrm{d}E \int \mathrm{d}r \, f(E) \Im \, \mathrm{Tr}[G^{e,e}(\vc{x}, \vc{x}; E)] + [1-f(E)] \Im \, \mathrm{Tr}[G^{h,h}(\vc{x}, \vc{x}; E)], \\
    \boldsymbol{\mu} &=& -\frac{1}{\pi} \int_{-\infty}^{\infty} \mathrm{d}E \int \mathrm{d}r \, f(E) \Im \, \mathrm{Tr}[\boldsymbol{\sigma} G^{e,e}(\vc{x}, \vc{x}; E)] + [1-f(E)] \Im \, \mathrm{Tr}[\boldsymbol{\sigma} G^{h,h}(\vc{x}, \vc{x}; E)], \\
    \chi &=& -\frac{1}{4\pi} \int_{-\infty}^{\infty} \mathrm{d}E \int \mathrm{d}r \, [1-2f(E)] \left( \Im \, \mathrm{Tr}[G^{e,h}(\vc{x}, \vc{x}; E)] + \Im \, \mathrm{Tr}[G^{h,e}(\vc{x}, \vc{x}; E)] \right).
\end{eqnarray}
Here, $f(E)$ refers to the Fermi-Dirac distribution and the trace is taken over the combined angular-momentum and spin index $L=(\ell,m,s)$, used in the expansion of the KKR Green's function around scattering sites~\cite{Ebert2011}. Note that this form of the anomalous density assumes $s$-wave pairing~\cite{Suvasini1993, Csire2015}. It is however possible to generalize the form of $\lambda$ and $\chi$ from simple real-valued scalars to more complex matrix forms in $L$. This approach can then be used to include a triplet order parameter in the self-consistent solution of the KS-BdG equation~\cite{Ghosh2020} during which, for a given fixed set of $\lambda_i$, $V_\mathrm{eff}$ and $\chi$ are converged which results in observables such as the superconducting gap in the band structure.

\subsection{Impurities and disorder in superconductors}
\label{sec:ImpEmbedding}

\begin{figure}
     \centering
    \includegraphics[width=0.9\textwidth, trim={0 -0.5cm 0 0},clip]{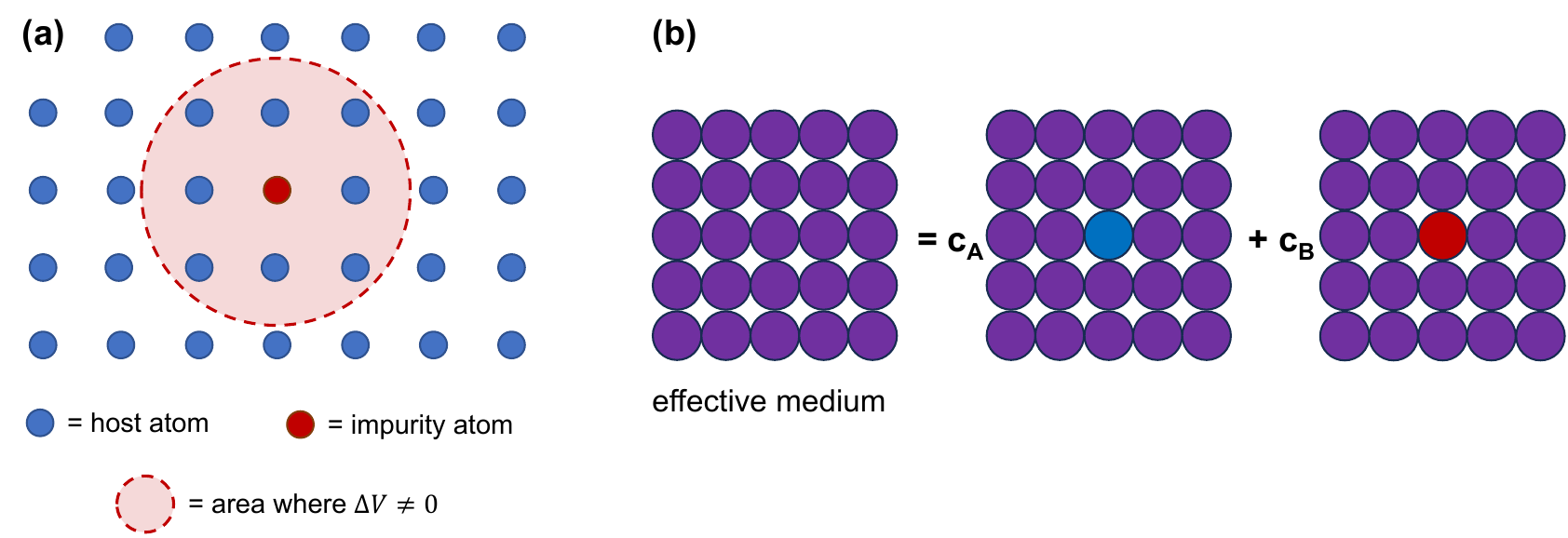}
    \caption{Impurity embedding and random disorder within the KKR method. \textbf{(a)} In the ab-initio impurity embedding, a defect introduces a change in the potential locally around the defect site. This is typically screened by the electrons of the host crystal so that the area where $\Delta V\ne 0$ is often short-ranged. \textbf{(b)} Effective medium of the Coherent Potential Approximation (CPA). The effective medium is constructed from the constituents of the random alloy in such a way that, on average, the individual components do not introduce any additional scattering.}
    \label{fig:impMethods}
\end{figure}

The Green's function formulation of KKR has the distinct advantage that impurities and disorder are easy to describe. This is due to the use of the Dyson equation that allows to connect the Green's function for a host crystal $G_\mathrm{host}$, which is routinely calculated for periodic systems with the KKR method, to the Green's function $G_\mathrm{imp}$ that describes an impurity embedded into the perfectly ordered host crystal. With $\Delta V = V_\mathrm{imp} - V_\mathrm{host}$, the impurity Dyson equation reads
\begin{equation}
    G_\mathrm{imp}(\vc{x}, \vc{x'}; E) = G_\mathrm{host}(\vc{x}, \vc{x'}; E) +  \int\mathrm{d}^3x\,G_\mathrm{host}(\vc{x}, \vc{x''}; E)\Delta V(\vc{x''})G_\mathrm{imp}(\vc{x''}, \vc{x'}; E).
\end{equation}
As in regular DFT for periodic systems, the impurity's charge density is related to the impurity potential via the Poisson equation. Thus, the impurity embedding requires a self-consistent solution of the impurity Dyson equation with updating $\Delta V$. 
However, since impurities are typically screened by the host's electrons, there is only a small region around the impurity where $\Delta V \ne 0$. 
Thus the impurity Dyson equation only needs to be solved in a small \emph{impurity cluster} around the defect atom, as illustrated in Fig.~\ref{fig:impMethods}(a).
This makes the Green's-function-based ab-initio impurity embedding very efficient, especially for the description of the dilute limit of defects. In contrast, wave-function-based DFT  requires periodically repeated supercells, that typically have to be large in order to avoid spurious interaction between periodic images of the impurities.

An alternative approach to the dilute limit treated within the ab-initio impurity embedding is to describe random disorder through alloying. Disorder at finite concentrations can be described within the Coherent Potential Approximation (CPA)~\cite{Faulkner,Ginatempo1988}. In this Green-function based method, an effective medium is defined by a local Green function and an averaged local scattering $\bar{t}$-matrix. These are found self-consistently by an implicit equation, where the known $t$-matrices of the alloy atoms enter, which are derived directly from the atom's potential. In contrast to the impurity embedding, which includes a cluster of neighboring atoms around the defect, the CPA is a single-site theory that is, however, known to work well in metallic alloys in three dimensions, as far as average quantities are concerned~\cite{Ebert2011}.

In this work, we employ both ab-initio impurity embedding \emph{and} CPA to describe transition metal doping in the $s$-wave superconductor Pb. The combination of the two approaches allows to study properties of single defects as well as the formation of impurity bands and the breakdown of superconductivity with increasing magnetic doping. Both approaches are formulated in the KS-BdG method where Green's functions, potentials for host and impurity atoms, as well as $t$-matrices entering the self-consistent equations of CPA contain particle and hole components leading to the four-component Green's function given in Eq.~\eqref{eq:GFBdG}.

\subsection{Computational details}
\label{sec:CompDetails}

\begin{figure}
    \centering
    \includegraphics[width=\linewidth]{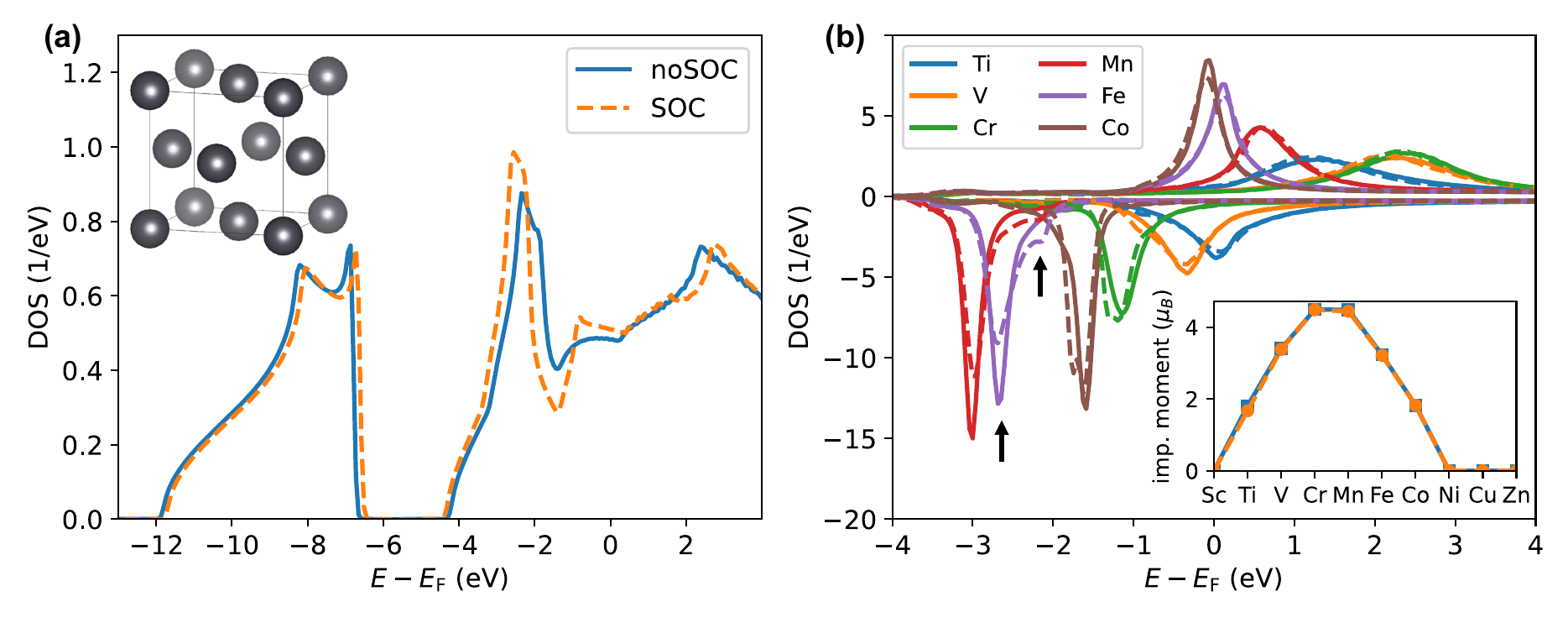}
    \caption{Normal state electronic structure of bulk Pb and embedded transition metal impurities. \textbf{(a)} Normal state density of states (DOS) of bulk Pb calculated with and without SOC. The inset shows the fcc crystal structure of Pb. \textbf{(b)} Impurity DOS of the magnetic $3d$ impurities. Full and broken lines indicate calculation results without and with SOC, respectively. SOC-induced splittings are highlighted by the black arrows for the Fe majority DOS. The inset in (b) shows the magnetic moment of the $3d$ impurities embedded as substitutional defects into Pb where the full blue (dashed orange) line refers to calculation results neglecting (including) SOC.}
    \label{fig:NormalState}
\end{figure}

Our calculations are performed with the JuKKR code~\cite{jukkr} with the help of the AiiDA-KKR interface~\cite{aiida-kkr-paper, aiida-kkr-code} to the AiiDA infrastructure~\cite{aiida}. We use the experimental lattice constant for bulk Pb of $4.95\,\mathrm{\AA}$ which crystallizes in the face centered cubic crystal structure~\cite{Bouad2003}, that is shown as an inset in Fig.~\ref{fig:NormalState}(a).
The transition metal impurities are placed in substitutional positions in impurity clusters that include the first shell of neighboring atoms, or within the (single-site) CPA. The superconducting state is calculated based on the solution of the Bogoliubov-de Gennes equation incorporated into the relativistic Korringa-Kohn-Rostoker Green function (KKR) method as described in detail in Ref.~\onlinecite{Ruessmann2022a}. Scalar relativistic corrections are included, and we perform two kinds of calculations either (i) including or (ii) neglecting the effects of spin-orbit coupling.

We use an angular momentum cutoff of $\ell_\mathrm{max} = 2$ and include corrections for the exact shape of the cells around the atoms (i.e.\ full-potential calculations)~\cite{Stefanou1990,Stefanou1991}. We use the local density approximation (LDA)~\cite{Vosko1980} for the parametrization of the normal state exchange correlation functional and the semi-phenomenological approach of Suvasini \textit{et al.}~\cite{Suvasini1993} for the superconducting coupling. The coupling constant $\lambda$ is set such that the superconducting gap of bulk Pb reproduces the experimental value of $\Delta \approx 1.4\,\mathrm{meV}$~\cite{Ruby2016, Khasanov2021}. This value of $\lambda$ is then kept constant in the subsequent CPA calculations that include transition metal impurities in Pb. Both in the calculations for impurity embedding and within the CPA, we set $\lambda$ to zero on the impurity site. Since the impurity concentration $x$ is low (typically $x<0.1$\%), a constant value of $\lambda$ independent of $x$ is a realistic assumption. Since we wish to describe small concentrations of magnetic impurities below the percolation limit, the paramagnetic state is a more realistic model than a ferromagnetic calculation. We follow the concept of the disordered local moments (DLM) within the CPA in order to describe the magnetically-disordered state~\cite{Pindor1983}. Thus, we define two species of magnetic defects with opposite magnetic moment orientations, e.g., Mn$^{\uparrow}$ and Mn$^{\downarrow}$. Then, the paramagnetic state Pb$_{1-x}$Mn$_x$ is calculated as an alloy of three components, Pb$_{1-x}$Mn$^{\uparrow}_{x/2}$Mn$^{\downarrow}_{x/2}$.

\section{RESULTS}
\label{sec:results}

\subsection{SOC-induced splitting of Yu-Shiba-Rusinov states}
\label{sec:YSR}

\begin{figure}
    \centering
    \includegraphics[width=\linewidth]{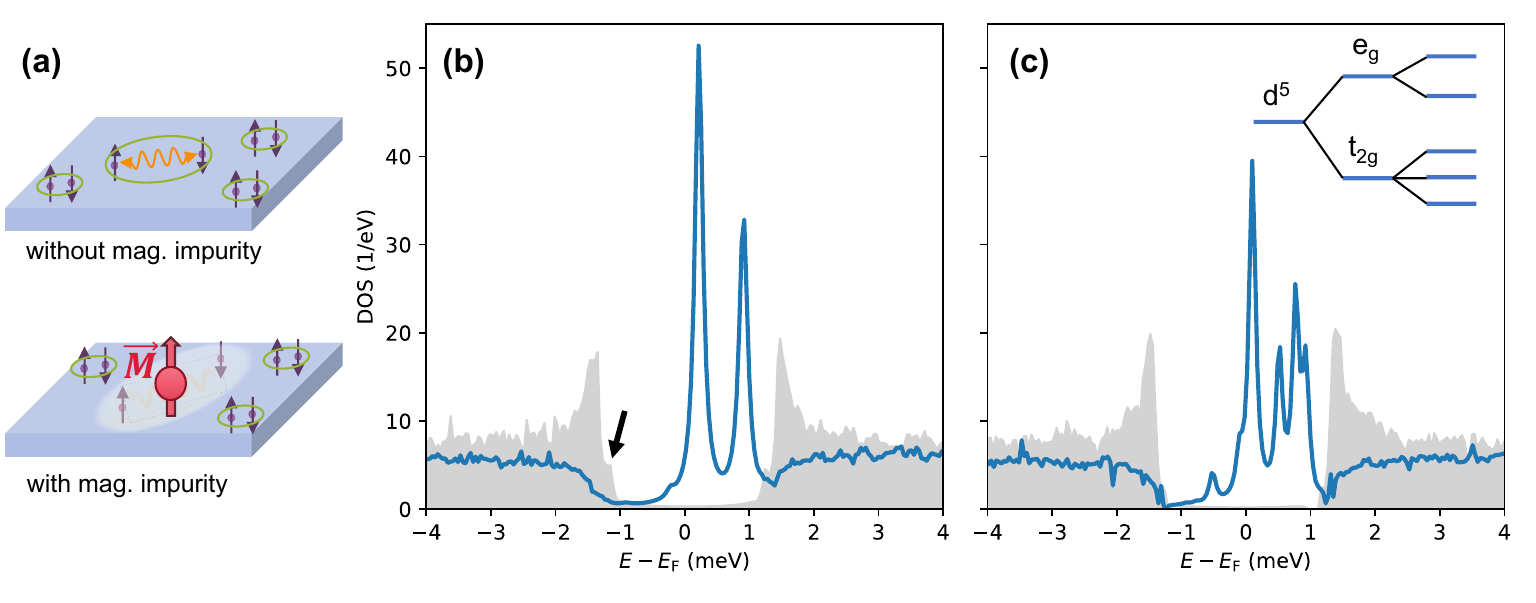}
    \caption{Yu-Shiba-Rusinov (YSR) states of Fe in Pb. \textbf{(a)} 
    Illustration of the pair breaking of singlet cooper pairs at magnetic defects. \textbf{(b,c)} Impurity DOS showing the YSR states of Fe without and with SOC included in the Pb host. The superconducting DOS of the Pb host crystal (scaled by a factor 20)  is given as the grey background. The arrow in (b) highlights the gap anisotropy (visible as shoulder due to a second coherence peak) that is more pronounced in Pb without SOC. The inset in (c) illustrates the crystal field and SOC induced splittings of the five $d$ orbitals in the fcc matrix of Pb.}
    \label{fig:YSR}
\end{figure}

Lead is a heavy metal where spin-orbit coupling (SOC) effects are particularly pronounced. This is visible in the density of states (DOS), that is shown with and without SOC in Fig.~\ref{fig:NormalState}(a). 
In the superconducting state, Pb is a two-band superconductor~\cite{Ruby2015, Gozlinski2023} due to different sheets that compose the Fermi surface~\cite{Floris2007, Tombulk}. This results in a gap anisotropy~\cite{Tombulk} that is seen as the shoulder at the edge of the coherence peak of the Pb host crystal (grey background spectra in Figs.~\ref{fig:YSR}(b)). With SOC, the signature of this anisotropy, however, reduces which may explain why some experiments are not able to observe the two-band superconductivity of Pb~\cite{Khasanov2021}, whereas it is visible in other experiments~\cite{Ruby2015, Gozlinski2023}.

We embedded the $3d$ series of transition metal (TM) atoms as substitutional defects in bulk Pb. The inset in Fig.~\ref{fig:NormalState}(b) shows the resulting impurity spin moment. We find that Ti, V, Cr, Mn, Fe and Co impurities ($Z=22$ to 27) develop a spontaneous magnetic moment that follows Hund's second rule with a maximum of the magnetic moment for half-filled $d$-shells. When the $d$-shell is unoccupied (Sc) or almost full (Ni, Cu and Zn), the impurities turn out to be non-magnetic. These trends are well known for transition metal impurities in non-magnetic metals with low electron density, and do not depend on whether we neglect or include SOC, as seen from the negligible differences in the impurity moments. However, SOC effects can be uncovered in the impurity DOS, shown in Fig.~\ref{fig:NormalState}(b). SOC induces splittings on top of the prevailing crystal field splitting in $e_g$ and $t_{2g}$ states, typically present in fcc crystal structures~\cite{Saunderson2022}. The SOC-induced splitting is particularly pronounced in the majority spin channel of Fe impurities, as highlighted by the black arrows. 
Moreover, it is well-known that SOC also induces orbital magnetic moments~\cite{Bruno1989}. We find sizable orbital moments of $0.14$, $0.02$, $-0.04$, $-0.13$, $-0.44$, and $-0.45$ for Ti, V, Cr, Mn, Fe and Co impurities, respectively. This trend follows the typical behavior for TM impurities in host materials with low electron density. The sign change from positive to negative values, which means a change from parallel to antiparallel orientation of spin and orbital magnetic moments, is consistent with Hund's third rule. These chemical trends are well known from transition metal impurities in other metallic host crystals~\cite{Podloucky1980, Dederichs1991, Popescu2001}.

In the superconducting state, the presence of a magnetic exchange field generally leads to breaking of singlet Cooper pairs~\cite{Balatsky2006}, as illustrated in Fig.~\ref{fig:YSR}(a). Cooper pairs in conventional $s$-wave superconductors form from a coherent superposition of two electrons of opposite spin. In the presence of a magnetic field, the exchange field of a magnetic impurity introduces different scattering potentials for the two electrons of the Cooper pair which results in a breaking of the coherent superposition. As a consequence, this leads to pairs of in-gap states within the superconducting gap of the host material known as Yu-Shiba-Rusinov (YSR) states~\cite{Yu1965, Shiba1968, Rusinov1969}. Here we calculate YSR states of magnetic TM impurities embedded in bulk Pb. For an Fe impurity the YSR states are shown in Fig.~\ref{fig:YSR}(b,c). Without SOC, two pronounced YSR states appear in the spin-integrated DOS at positive energies. In a simple model that assumes isotropic scattering, the energy of YSR states is given as~\cite{vonOppenFranke2021, Saunderson2022}
\begin{equation}
    \varepsilon = \pm \Delta_0 \frac{1-\alpha^2+\beta^2}{\sqrt{(1-\alpha^2+\beta^2)^2+4\alpha^2}},
    \label{eq:YSR}
\end{equation}
where $\alpha=\pi N_0 JS$ is the pair-breaking part of the scattering potential ($N_0$ is the superconductor's normal DOS at $E_\mathrm{F}$ and $JS$ is the exchange energy a classical spin $S$ experiences) and $\beta=\pi N_0 V$ is related to the non-magnetic part of the impurity's scattering potential $V$. From Eq.~\eqref{eq:YSR} it is obvious that YSR states come in pairs (leading $\pm$ on the r.h.s.), which are symmetric with respect to flipping the electron's spin. Their intensities are however generally not identical due to the difference in the impurity DOS between minority and majority spin channels. In our spin-integrated DOS shown in Fig.~\ref{fig:YSR}(b), the majority spin-channel produces YSR states that are almost two orders of magnitude smaller than the minority spin channel. The splitting into two peaks can be attributed to the crystal field splitting the Fe impurity experiences in the fcc matrix of the Pb host crystal. This is understood from the different exchange field an electron coupling to the $e_g$ or $t_{2g}$ manifold of $d$-states experiences~\cite{Saunderson2022}. The crystal field splitting enters Eq.~\eqref{eq:YSR} via different values for $JS$ and thus different values of $\alpha$ for the ${e_{g}}$ and $t_{2g}$ states.

\begin{figure}
    \centering
    \includegraphics[width=0.7\linewidth]{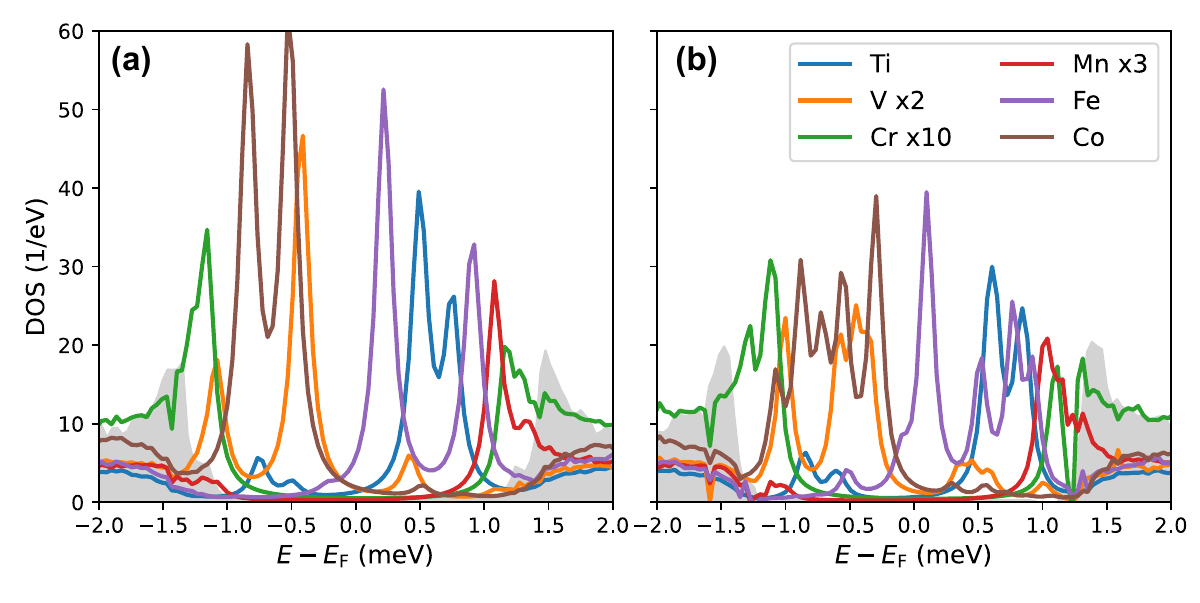}
    \caption{YSR states of different magnetic $3d$ transition metal impurities in bulk Pb. Spin-integrated impurity DOS showing YSR states calculated \textbf{(a)} without SOC and \textbf{(b)} including SOC. The grey-shaded background shows the host's DOS (scaled up by a factor 20) where the superconducting gap in the host crystal is visible.}
    \label{fig:YSR2}
\end{figure}

The SOC-induced splitting of the majority states of Fe also leads to a changing energy difference between $e_g$ and $t_{2g}$ states. We attribute the appearance of a YSR state closer to $E=E_\mathrm{F}$, when SOC is included (\textit{cf.}\ Fig.~\ref{fig:YSR}(c)), to this fact. In addition to this global shift, we also find that the $e_g$ and $t_{2g}$ manifolds further split, leading to the complex multi-peak structure in the YSR spectrum of Fe. A comparison of the YSR states of different magnetic TM impurities is shown in Fig.~\ref{fig:YSR2}. The differences in the YSR spectra demonstrate the chemical trends for changing magnetic impurity. From the impurity DOS calculation where SOC is neglected, shown in Fig.~\ref{fig:YSR2}(a), we deduce that the size of the crystal field splitting depends on the impurity type. We find the largest values for V and Fe impurities and smaller splittings for Co, Ti, Mn and Cr (ordered by the size of the splitting from highest to lowest). The filling of the impurity's $d$-shell, that is related to its magnetic moment, also strongly influences where the YSR states reside inside the host's superconducting gap. Whereas we find that the dominant YSR peaks arise at positive energies for Ti, Mn and Fe, we find the opposite behavior for V, Cr and Co. Furthermore, for Cr and Mn the YSR states stick to the edge of the gap but for V, Fe and Co we find them closer to the center of the gap. The SOC-induced splitting on top of the crystal field splitting of YSR states can furthermore lead to a significant broadening of the YSR spectrum as seen in Fig.~\ref{fig:YSR2}(b). This is particularly pronounced for Co where we see that the YSR spectrum fill almost the entire lower half (i.e.\ at negative energies) of the superconducting gap. 

Previous calculations for bulk Pb, that however neglected the effect of SOC, agree well with our results without SOC~\cite{Saunderson2022}.
Measurements by Ruby \textit{et al.}~\cite{Ruby2016} of Mn impurities on Pb(110) confirm that the main spectroscopic signature is found close to the coherence peak of Pb. This was recently also found theoretically by Wu \textit{et al.}~\cite{Wu2023}, where it was pointed out that the exact position of YSR peaks at the surface depends sensitively on relaxations of the distance of the impurity to the surface. However, in the bulk, where we place our impurities, the impurities experience a higher coordination number of neighboring host atoms which will generally reduce the effects of lattice relaxations.  
The chemical trends that we observe for different TM impurities in Pb are, furthermore, consistent with STM experiments done for different magnetic impurities on the $s$-wave superconductor Nb~\cite{Kuster2021}. For magnetic impurities on Nb it was further discussed that the direction of the magnetic of Mn impurities on the (110) surface and the size of SOC have a strong influence on the location of YSR peaks within the superconducting gap~\cite{Laszloffy2023}. We also expect these effects to be reduced for impurities in the highly symmetric bulk fcc lattice.

\subsection{Breakdown of superconductivity at finite impurity concentrations}
\label{sec:CPA}

\begin{figure}
    \centering
    \includegraphics[width=0.8\linewidth]{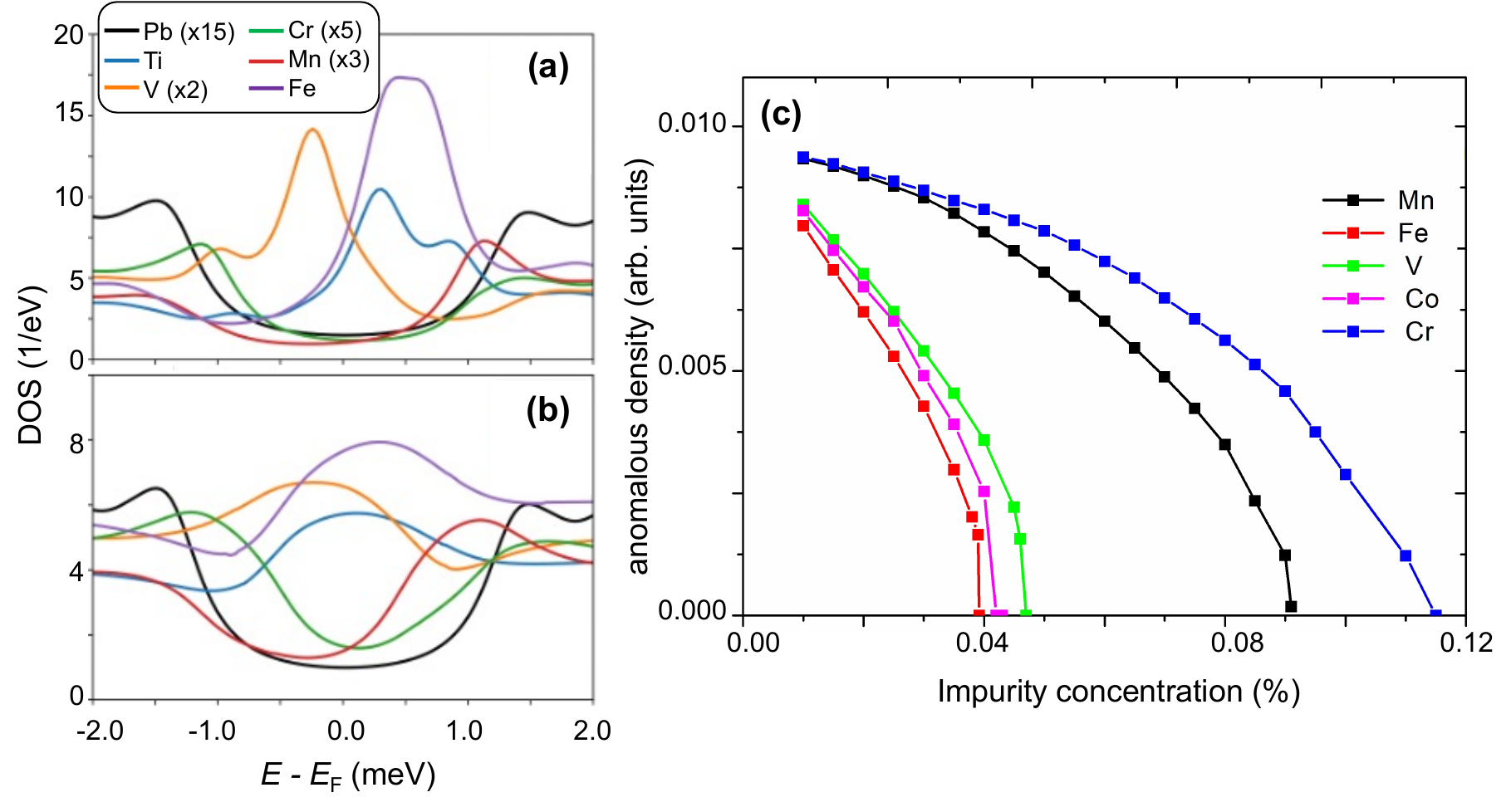}
    \caption{Random alloying of magnetic impurities into superconducting Pb, computed with the CPA method. \textbf{(a,b)} Broadening of the YSR peaks for different magnetic impurities at concentrations of $1\,\mathrm{ppm}$ and $100\,\mathrm{ppm}$. \textbf{(c)} Breakdown of superconductivity with increasing concentration of magnetic defects embedded into Pb. Shown is the anomalous density, which is the order parameter for superconductivity in the KS-BdG method.}
    \label{fig:CPA}
\end{figure}

In the dilute limit, impurity states are localized around isolated impurity atoms and YSR bound states appear as discussed in the previous section. With increasing concentration, however, these YSR states will overlap and impurity bands can gradually form that might have a profound effect on the electronic structure. Analogous to critical magnetic fields where superconductivity breaks down, at a critical concentration of magnetic impurities mixed into an $s$-wave superconductor, a breakdown of superconductivity is expected~\cite{Balatsky2006}. We study this effect within the CPA, where we simulate the superconducting state of Pb with increasing concentrations of magnetic defects. The results are summarized in Fig.~\ref{fig:CPA}. For this study we chose to neglect SOC effects in these calculations because we expect the disorder broadening to be more significant than the SOC-induced splitting seen for the YSR states of some TM impurities. The CPA impurity DOS is shown in Fig.~\ref{fig:CPA}(a,b) for impurity concentrations of $1\,\mathrm{ppm}$ and $100\,\mathrm{ppm}$ ($0.01\%$). Indeed, we find a strong broadening of the YSR states that prevents the clear identification of distinct $e_g$ and $t_{2g}$ states for concentrations as low as $100\,\mathrm{ppm}$, which is well below the critical concentration where superconductivity breaks down. Overall we find that the CPA results of the impurity DOS agree well with the calculations based on ab-initio impurity embedding, \textit{cf.}\ Fig.~\ref{fig:YSR2}(a). We note that the formation of bonding/anti-bonding pairs when YSR states of close-by atoms overlap~\cite{Flatte2000}, can further shift the location of YSR states. This effect was, for example, discussed in pairs of Mn atoms on the Nb(110) surface~\cite{Nyari2021} and it is also included in our results within the CPA. The disorder broadening, however, prevents the observation of distinct shifts. 

Figure~\ref{fig:CPA}(c) shows the breakdown of superconductivity with increasing impurity concentrations for different magnetic TM impurities. This is shown in terms of the integrated anomalous density $\chi$. 
In the weak coupling limit of the BCS theory~\cite{BCS}, it is known that $T_c$ is proportional to the superconducting gap $\Delta$, which in turn is proportional to $\chi$ in our formalism. At $x=0$ we obtain a maximum $\chi=\chi_0$, which vanishes for $x\geq x_c$. It is far more practical to monitor the superconducting state through the anomalous density than through the superconducting gap, because the latter is blurred by the presence of in-gap YSR impurity states and because close to $x_c$ we expect gapless superconductivity. In all cases, $\chi$ decreases as a function of the concentration up to a critical point $x_c$, after which it vanishes. As far as can be inferred from our numerical results, the curve $\chi(x)$ is continuous. At $x\lessapprox x_c$ it decays linearly and shows a discontinuity of the first derivative at $x_c$. This behaviour is well-known from the Abrikosov-Gor'kov theory~\cite{AG} and it is, for example, seen in conventional magnetically-doped superconductors~\cite{WOLOWIEC2015113}.

We find large differences in the values of $x_c$ for different magnetic impurities. These values, however, do not seem to correlate clearly with the values of the magnetic moments (\textit{cf.}\ Fig.~\ref{fig:NormalState}). In principle, one would expect higher impurity moments to suppress the superconducting state at lower concentrations, since the spin-polarised part of the potential, giving rise to the magnetic moment, is also the cause of Cooper-pair-breaking. In our calculations we find that $x_c$ is low for Ti and V, then grows for Cr and Mn, where the magnetic moments are more sizeable, and then falls off again for Fe and Co. We suspect that this non-linear behaviour is related to the existence of $d$ level resonances at the Fermi energy, which is captured in our DFT calculation. As shown in Fig.~\ref{fig:NormalState}(b), Fe and Co have a minority-spin resonance peaked at $E_\mathrm{F}$, while V has a majority-spin resonance peaked at $E_\mathrm{F}$. These impurities exhibit a low critical concentration, between $0.04\%$ and $0.05\%$. For Mn and Cr, on the other hand, $E_\mathrm{F}$ lies between the majority-spin and minority-spin resonances. Between the two, the Mn minority-spin peak is slightly closer to $E_\mathrm{F}$ than the Cr majority-spin peak. This is reflected in the values of $x_c=0.091\%$ for Mn and $0.115\%$ for Cr. Moreover, we find a correlation between the critical concentration and location of the YSR peaks within the superconducting gap of the host. For Mn and Cr, the YSR states stick to the edge of the superconducting gap whereas we find that they populate the central region for Fe, V and Co, as seen in Figs.~\ref{fig:YSR2}(a) and \ref{fig:CPA}(a,b). 

\section{DISCUSSION AND PERSPECTIVE}
\label{sec:discussion}

Interfacing superconducting and magnetic materials forms the basis for the emergent field of superconducting spintronics~\cite{Eschrig2015, Linder2015, Cai2023, Melnikov2022}. While conventional $s$-wave superconductivity breaks down at critical magnetic fields, proximity between magnetic and superconducting materials provides a fruitful basis for unconventional superconductivity. In material systems where the macroscopically coherent superconducting states coexist with microscopic exchange interaction at magnetic atoms or layers, for example,  triplet superconductivity, odd-frequency pairing, long-range equal-spin supercurrents or Majorana qubits can be engineered~\cite{Eschrig2015, Cai2023}. Devices that combine superconducting and magnetic materials allow to realize unconventional transport with superconducting spin valves or unconventional Josephson junctions consisting of superconductor ferromagnet (S/F) interfaces and S/F/S heterostructures~\cite{Melnikov2022}. 

\begin{figure}
    \centering
    \includegraphics[width=0.5\linewidth, trim={0 -0.5cm 0 0},clip]{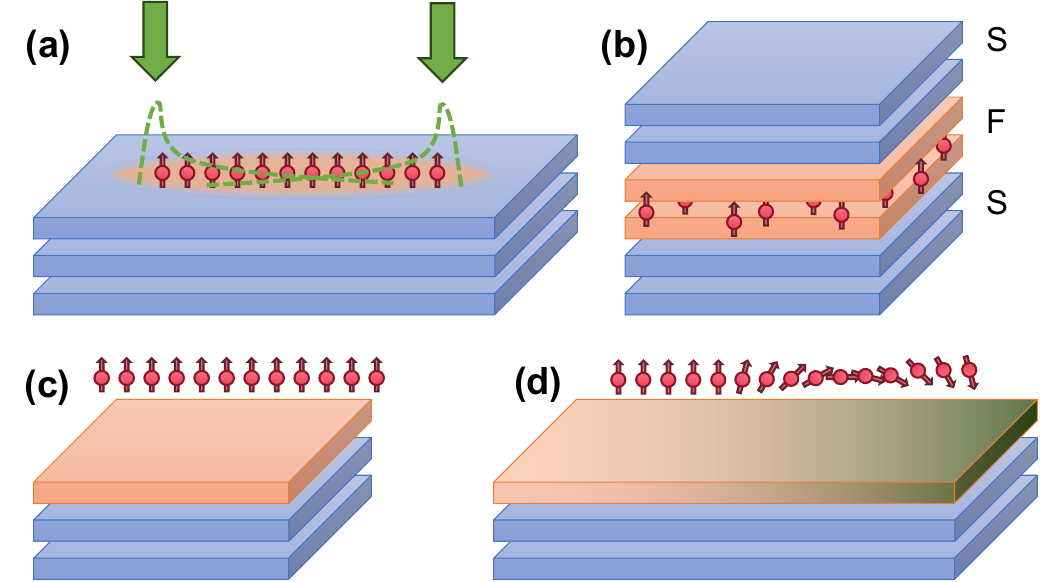}
    \caption{Possible geometries where superconductivity and magnetism come together. Layers of atoms are illustrated by filled rectangles where the superconductor is coloured in blue and (ferro)-magnetic layers are coloured in orange. \textbf{(a)} Chain of magnetic atoms on a superconducting surface where Majorana zero modes may appear at the ends of the chain. \textbf{(b,c)} Interfaces between superconductors (S) and ferromagnets (F) in form of a S/F/S junction or S/F interface. \textbf{(c)} Non-collinear magnetic layer in contact to a superconductor. The magnetic ordering of the magnetic layers is (c,d) is illustrated by the direction of the spins in the chain above the surface.}
    \label{fig:SFS}
\end{figure}

An example are materials that provide large interfacial SOC in superconducting junctions, which allows to realize devices that show a large magnetoresistance effect~\cite{Martinez2020}.
Interesting effects arise not only at isolated magnetic defects in superconductors, which is discussed in this work in Sec.~\ref{sec:results}, but in a variety of possible S/F geometries, \textit{cf.}\ Fig.~\ref{fig:SFS}. For example, chains of magnetic atoms in proximity to a superconductor can, together with SOC,  lead to the formation of Majorana zero modes at the end of the chains~\cite{Kitaev2001, NadjPerge2014}. With distributed magnetic doping or entire magnetic layers, S/F interfaces and S/F/S heterostructures, illustrated in Fig.~\ref{fig:SFS}(b,c), can be engineered. Finally, non-uniform magnetic order can have a remarkable effect in S/F hybrid structures. For instance, the existence of domain walls in magnetic layers can stabilize superconductivity~\cite{Yang2004}. Moreover, topological superconductivity together with spin-triplet correlations is proposed to exist in skyrmion lattices in contact to superconductors~\cite{Mascot2021}.

In F/S and S/F/S heterostructures, the proximity effect  will in general depend on material properties such as crystal symmetries, orbital character of the electronic structure of its constituents. This calls for a microscopic modelling of the interfaces that include all these effects. The KS-BdG method is ideally suited for this task. It naturally accounts for crystal symmetries and has chemical accuracy, taking into account different orbital character of a material's electrons. Different geometries from single impurities over nanoclusters to interfaces and heterostructures including SOC and non-collinear magnetic textures are routinely described with DFT methods. 
The example of TM impurities embedded into the $s$-wave superconductor Pb, which we discussed in Sec.~\ref{sec:results}, demonstrates the power of the KS-BdG approach. Since this method is based on DFT, multi-band effects are automatically included and material-specific predictions are possible, even at the sub-meV energy scale necessary in the field of superconductivity. This is illustrated by the splitting of the manifold of YSR states that is derived from a combination of crystal field and SOC splittings of $d$-states. The chemical trends uncovered in the comparison of different transition metal impurities can have a profound effect on the details of the YSR in-gap states. This results in vastly different YSR peak positions and different critical concentrations, where superconductivity breaks down, for different transition metal impurities. 

Applying the KS-BdG method to diverse material classes where superconductivity, SOC, and magnetism can coexist offers enticing research opportunities. For example, layered van der Waals (vdW) materials offer rich physical properties that can be combined easily in heterostructes~\cite{Geim2013}. A prominent vdW superconductor is NbSe$_2$~\cite{Wickramaratne2020}, where, for instance, YSR states around magnetic Fe impurities are found to extend over several nm~\cite{Menard2015}. 
NbSe$_2$ also shows signatures of equal-spin triplet pairs under an applied magnetic field, which are of great interest for superconducting spintronics and topological superconductivity~\cite{Kuzmanovic2022}.
Furthermore, a field-free version of a Josephson diode was recently realized in a heterostructure of NbSe$_2$ with Nb$_3$Br$_8$~\cite{Wu2022}. This accomplishes the superconducting diode effect with a built-in magnetic field, that was originally found in Nb/Ta/V multilayers with an applied \emph{external} field~\cite{Ando2020}.

With regard to the field of superconducting spintronics harnessing and extending the novel capabilities of the KS-BdG method shows great promise. For instance, it is possible to predict unconventional pairing such as inter-orbital Cooper pairs that have singlet-triplet mixing~\cite{Bahari2022, Ruessmann2023}. Signatures of triplet pairing, which is a hallmark for unconventional superconductivity, are also accessible in the KS-BdG method. The output anomalous density matrix $\chi$ (introduced in Sec.~\ref{sec:methods}) in general has both singlet and triplet components. During the self-consistent solution of the KS-BdG equation, in addition to (or even instead of) the singlet component alone, a generalized set of coupling constants can be included. This was, for example, done to study triplet pairing in the unconventional superconductor LaNiGa$_2$~\cite{Ghosh2020}. Investigating triplet order in the superconductivity of S/F interfaces and S/F/S junctions from the KS-BdG method will be of interest for the materials optimization challenges in superconducting spintronics. Moreover, being able to predict transport properties from microscopic band structure simulations of the KS-BdG calculations would also be of great interest. This would allow to connect the microscopic KS-BdG picture to the realm of mesoscopic physics where, for instance, transport characteristics of Josephson junctions are studied. A possible avenue towards that goal might be the pursuit of DFT-based simulations of the supercurrent, that was recently done in the ballistic limit of magnetic Josephson junctions consisting of Nb/Ni/Nb~\cite{Ness2022}.

In summary, the KS-BdG method offers unique insights into the electronic structure, proximity effects and superconducting order based on the microscopic atomic structure. Applying this methodology to S/F/S-type heterostructures is expected to give valuable insights in the future, tackling challenging materials problem on the way towards applications in superconducting spintronics and quantum information technology.

\section*{ACKNOWLEDGMENTS}
    We thank the Bavarian Ministry of Economic Affairs, Regional Development and Energy for financial support within the High-Tech Agenda Project ``Bausteine für das Quantencomputing auf Basis topologischer Materialien mit experimentellen und theoretischen Ansätzen''.
	Furthermore, this work was funded by the Deutsche Forschungsgemeinschaft (DFG, German Research Foundation) under Germany's Excellence Strategy -- Cluster of Excellence ``Matter and Light for Quantum Computing'' (ML4Q) EXC 2004/1 -- 390534769.
	Part of this work has been performed under the Project HPC-EUROPA3 (INFRAIA-2016-1-730897), with the support of the EC Research Innovation Action under the H2020 Programme.
	We are also grateful for computing time granted by the JARA Vergabegremium and provided on the JARA Partition part of the supercomputer CLAIX at RWTH Aachen University (project ``jara0191'') and of the supercomputer JURECA~\cite{jureca} at Forschungszentrum Jülich (project ``superint''), and this work was supported by computational time granted from the Greek Research \& Technology Network (GR-NET) in the National HPC facility---ARIS---under projects ID 2DMATFUN and TopMagX3.

\bibliography{references} 
\bibliographystyle{spiebib} 

\end{document}